\documentclass[a4paper,11pt]{article}
\usepackage{pos}

\def\d{\delta}

\def\l{\lambda}
\def\m{\mu}
\def\n{\nu}

\def\p{\pi}                     
\def\s{\sigma}                  

\def\x{\xi}

\def\L{\Lambda}

\def\ie{\mbox{\it i.e.}}

\def\svev#1{\left\langle #1\right\rangle}       

\def\bibi{\bibitem}
\def\ttl#1{{\it #1}}

\def\PiT{\Pi^{(1)}}
\def\hPiT{\hat\Pi}
\def\PiL{\Pi^{(0)}}
\def\Pidiff{\Pi^{(1-0)}}

\def\SNLO{S_{\rm NLO}}
\def\SHC{S_{\rm HC}}
\def\Dmix{\Delta_{\rm mix}}

\title{$S$ parameter from a prototype composite-Higgs model}

\author*[a]{Yigal Shamir}
\author[b]{Maarten Golterman}
\author[c]{William~I.~Jay}
\author[d]{Ethan T.~Neil}
\author[a]{Benjamin Svetitsky}

\affiliation[a]{Raymond and Beverly Sackler School of Physics and Astronomy,
  Tel~Aviv University,\\ 69978 Tel Aviv, Israel}

\affiliation[b]{Department of Physics and Astronomy,
  San Francisco State University,\\ San Francisco, CA 94132, USA}

\affiliation[c]{Theoretical Physics Department,
  Fermi National Accelerator Laboratory,\\ Batavia, Illinois, 60510, USA}

\affiliation[d]{Department of Physics, University of Colorado,\\
  Boulder, CO 80309, USA}

\emailAdd{shamir@tauex.tau.ac.il}

\abstract{We have calculated the low-energy constant $L_{10}$
in a prototype composite Higgs model with dynamical fermions
in two different representations of the gauge group.
The resulting contribution of the new strong sector
to the $S$ parameter is consistent with current bounds on the
vacuum misalignment parameter.  We end with a brief discussion
of future directions.}

\FullConference{%
 The 38th International Symposium on Lattice Field Theory, LATTICE2021
  26th-30th July, 2021
  Zoom/Gather@Massachusetts Institute of Technology
}

\begin{document}
\maketitle

\section{Introduction}
According to the composite Higgs paradigm, the Higgs boson is a
pseudo Nambu-Goldstone boson of a new strong interaction \cite{GK,DGK}.
This protects its mass from receiving large radiative corrections.
Hypercolor---the new strong interaction---may be operative at the few TeV scale
(see Refs.~\cite{RC,BCS,PW} for reviews).
One can also use the hypercolor interaction to generate
a partially-composite top quark via linear coupling
to a hypercolor baryon \cite{DBK}.

We have been studying a prototype composite-Higgs model
using lattice techniques \cite{meson,baryon,TACO1802,TACO1812,CLR2,ourL10}.
The model is an SU(4) gauge theory with 2 Dirac fermions in the
fundamental representation, together with 2 Dirac fermions
in the sextet, or two-index antisymmetric, representation.
Since the sextet is real, this is equivalent to 4 Majorana fermions.
Our work was the first lattice calculation
with dynamical fermions in two different representations.
For similar lattice calculations, see Refs.~\cite{CDPP,BP}.
For lattice work related to composite-Higgs models
based on an Sp(4) gauge theory, see Refs.~\cite{Sp4a,Sp4b,Sp4c,Sp4d}.

In itself, the fermion content of the prototype model is not enough
to meet phenomenological requirements.  But it is quite close to one of the
composite-Higgs models proposed by Ferretti and Karateev \cite{FerK},
labeled M6 in Ref.~\cite{diboson}, which contains 3 fundamental Dirac fermions
and 5 sextet Majorana fermions (see also Refs.~\cite{ferretti14,ferretti16}).
Note that the Ferretti-Karateev models are partially, but not
fully, ultraviolet complete: the origin of the 4-fermion couplings needed
to generate a partially composite top quark remains unspecified.
For a proposal for a full ultraviolet completion, see Ref.~\cite{CVZ}.

Here we report on our calculation of the low-energy constant $L_{10}$
\cite{ourL10},
which largely follows the QCD calculations of Refs.~\cite{JLQCD,DWF}.
We have calculated $L_{10}$ in the sextet sector
of our two-representation theory,
since in the M6 model the electro-weak symmetries
are embedded in the unbroken global symmetry of that sector.
Our result for $L_{10}$ allows us to constrain the contribution of
the hypercolor theory to the $S$ parameter in terms of the
vacuum misalignment parameter \cite{RC,BCS,PW}.

\section{Chiral perturbation theory}
In the continuum,
the correlator of a left-handed and a right-handed current is decomposed as
\begin{equation}
\label{JLJR}
\svev{J_{L\m}J_{R\n}} = (q^2 \d_{\m\n} - q_\m q_\n) \PiT(q^2)
+ q_\m q_\n \PiL(q^2) \ ,
\end{equation}
with transverse function $\PiT$ and longitudinal function $\PiL$.
The transverse function is given by
\begin{equation}
\label{PiT}
\PiT = 
\frac{F^2}{q^2} + \hPiT(q^2) \ ,
\end{equation}
where the leading-order pole reflects a kinematical singularity.
To next-to-leading order (NLO) in chiral perturbation theory
and for $N$ Majorana fermions \cite{BL,tworeps}
\begin{equation}
\label{pihat}
\hPiT(q^2) = 
\frac{N+2}{96\pi^2}
\left[\frac13+\log\left(\frac{M^2}{\mu^2}\right)-H(s)\right] + 8 L_{10} \ ,
\end{equation}
which includes the dependence on $L_{10}$.
Here $M$ is the pion mass and $\m$ is the renormalization scale.
The momentum dependence enters via
\begin{equation}
\label{Hs}
  H(s)=2s^2+s^3\log\left(\frac{s-1}{s+1}\right)\ ,
\end{equation}
with $s=\sqrt{1+4M^2/q^2}$.  The function $H(s)$ has no free parameters.
We will also need the difference
\begin{equation}
\label{pidiff}
\Pidiff = \PiT - \PiL \ = \ \frac{F^2}{q^2+M^2}+\hPiT(q^2) \ .
\end{equation}

Turning to the lattice calculation, our ensembles were generated with
dynamical Wilson-clover fermions for both representations \cite{meson}.
nHYP smearing was used in the fermion action \cite{HK,HHS},
and an NDS term was added to the gauge action to further improve
the performance of the smeared links \cite{NDS}.
In view of the importance of chiral symmetry for the calculation of $L_{10}$
we used staggered valence fermions, \ie, we did a mixed-action calculation.
$J_{L\m}$ and $J_{R\n}$ were constructed from the standard vector and axial
staggered currents, except again with nHYP links.
We made the usual replacement $q_\m \to \hat{q}_\m=(2/a)\sin(aq_\m/2)$,
with $a$ the lattice spacing.  The renormalization scale was $1/\m^2=t_0$,
where $t_0$ is the gradient-flow scale.

The pole terms, \ie, the first term on the right-hand sides
of Eqs.~(\ref{PiT}) and~(\ref{pidiff}), involve the pure valence pion,
so that $M=M_{vv}$ and $F=F_{vv}$.  At NLO,
the pion in the loop is a mixed sea-valence pion, hence, in Eq.~(\ref{pihat}),
as well as inside the argument $s$ of $H(s)$, $M=M_{vs}$, with
\begin{equation}
\label{Mvs}
M_{vs}^2 = \frac{M_{ss}^2 + M_{vv}^2}{2} + \frac{a^2}{t_0^2}\,\Dmix\ .
\end{equation}
Here $\Dmix\ge0$ is a new parameter peculiar to the mixed-action setting
\cite{BRS,BBRS,Dmix,BGS}.
In order to explore the possible effects of the next order
in chiral perturbation theory, we have augmented Eq.~(\ref{pihat})
by the (NNLO) analytic terms,
\begin{equation}
\label{NNLO}
t_0 \left(b_q\, q^2 + b_{ss}\, M_{ss}^2 + b_{vv}\, M_{vv}^2\right)
+ b_a\, \frac{a^2}{t_0} \ .
\end{equation}

\section{Results}
We used twelve $16^3\times32$ ensembles and three $24^3\times48$ ensembles
taken from Ref.~\cite{meson}, with gradient-flow scale $t_0/a^2$
in the range 0.9--2.7, and sea-pion mass $\sqrt{t_0} M_{ss}$
in the range 0.2--0.58.  After some exploratory studies, we settled on
a calculation with 7 valence masses $am_v$ in the range 0.01--0.05.
We successfully fitted $\Pidiff$, whose correlations turned out
to be smaller than those of $\PiT$, always including $L_{10}$ and $b_{vv}$
in the fit. In order to limit other sources of higher-order corrections
apart from the valence mass, we evaluated the correlators only at
the smallest time-like momentum.
We tried all 16 combinations of the 4 remaining parameters
($\Dmix$, $b_q$, $b_a$ and $b_{ss}$), always obtaining a good $p$-value.

\begin{figure}[t]
\begin{center}
\vspace*{-1ex}
\includegraphics[width=.9\columnwidth]{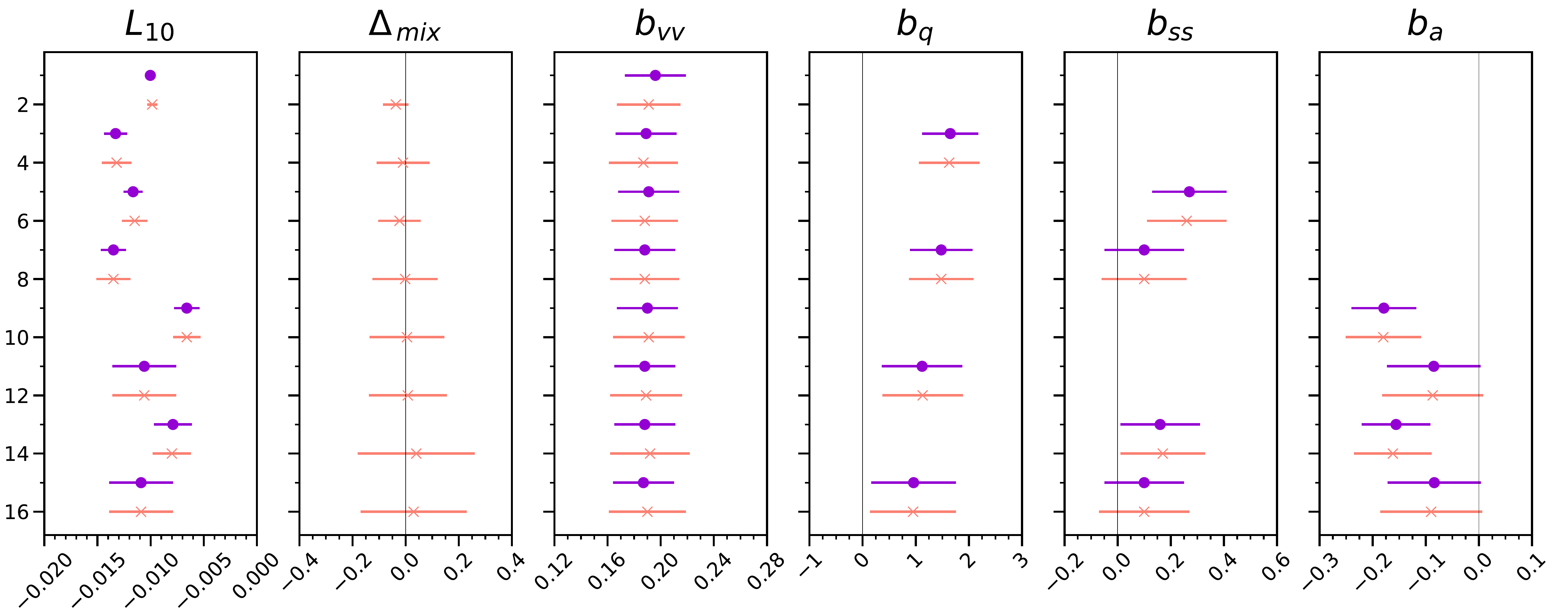}
\end{center}
\vspace*{-2ex}
\caption{
\label{stabplot}
Sixteen fits of $\Pidiff$ to data from all 7 valence masses.
All fits include $L_{10}$ and $b_{vv}$ as parameters
but have different combinations of $\Dmix$
and the other NNLO parameters $b_q$, $b_{ss}$ and $b_a$.
Fits without $\Dmix$ are shown in purple, and with $\Dmix$ in orange.}
\end{figure}

The results are shown in Fig.~\ref{stabplot}.
The NLO mixed-action parameter $\Dmix$ was always consistent with zero,
so we focus on the fits without $\Dmix$.
The value of $b_{vv}$ was stable and more than $5\s$ away from zero,
which explains why this NNLO parameter had to be included in all the fits.
Fit No.\ 1 includes only $L_{10}$ and $b_{vv}$,
and has a very small statistical error (about the size of the data point).
The remaining 7 fits with $\Dmix=0$ include different subsets
of the other NNLO parameters $b_q$, $b_a$ and $b_{ss}$.
They give rise to considerable variation in the value of $L_{10}$,
as well as to increasing statistical error.  The results of
fits 3, 5, and 9, which include only a single additional NNLO parameter
suggest that our main source of uncertainty is systematic.
We used the central values of these fits to bracket the systematic error
of our final result, and their statistical errors to estimate the
statistical error of this result, obtaining
\begin{equation}
\label{final}
  L_{10} = -0.0100(12)_{\rm stat}(35)_{\rm syst} \ .
\end{equation}
The central value of our final result coincides with that of
fit 1, which contains only $L_{10}$ and $b_{vv}$.
Changing the renormalization scale $\m$ in Eq.~(\ref{pihat})
from $1/\sqrt{t_0}$ to the sextet vector meson mass shifts the central value
of $L_{10}$ by about $-0.00035$, a 3.5\% shift.

The contribution of the hypercolor sector to the $S$ parameter is
\begin{equation}
\label{SHC}
\SHC = \x \SNLO \ .
\end{equation}
Here $\x = 2v^2/F_6^2$ is the vacuum misalignment parameter,
where $v=246$~GeV is the vacuum expectation value of the Higgs field
in the Standard Model, and $F_6$ is the decay constant
of Nambu-Goldstone bosons made of the sextet fermions in the chiral limit.
$\SNLO$ is given by \cite{PT,BL}
\begin{equation}
\label{SNLO}
\SNLO = -2\p \lim_{q^2\to 0} \hPiT(q^2)
= -\frac{N+2}{24\p}
  \left(1 + \log\left(\frac{M^2}{\m^2}\right) \right) - 16\p L_{10}  \ .
\end{equation}
Assuming our result for $L_{10}$ can be used in the M6 model
(for which $N=5$ in the above equation), we find for that model
\begin{equation}
\label{Sresult}
\SNLO = 0.8(2) \ ,
\end{equation}
where the error is dominated by the systematic error of $L_{10}$.
In order to arrive at this result we have assumed for simplicity
that the 14 pseudo Nambu-Goldstone bosons of the SU(5)/SO(5) coset
have a common mass $M$ about the size of the Higgs mass or somewhat larger.
The uncertainty due to the actual spread of (non-degenerate) masses
is expected to be much smaller than the already quoted error.
Also, we assumed $F_6 = 1.1$ TeV, the lowest value consistent with
the commonly quoted upper bound $\x \le 0.1$ \cite{RC,BCS,PW}.
The current experimental estimate is $S = -0.01(10)$, which implies
a $1\s$ upper bound of 0.09.  This yields an independent $1\s$ bound
\begin{equation}
\label{xbound}
  \x \le \frac{0.09}{0.8(2)} = 0.11(3) \ ,
\end{equation}
which is compatible with the upper bound mentioned above.


\section{Future prospects}
We previously reported an acute problem with our prototype
composite-Higgs model \cite{TACO1812,BQS19}.  The problem has to do with the
partially-composite top, which should receive its mass via
direct coupling to a hypercolor baryon with the same
Standard-Model quantum numbers as the top quark.
This coupling is facilitated by a 4-fermion operator, schematically,
$t Q qq$, where $t$ is the top quark, $Q$ is a sextet hypercolor fermion,
while $qq$ is a pair of fundamental hypercolor fermions.
The hypercolor singlet operator $Qqq$ generates a ``chimera'' baryon.

The 4-fermion coupling $G/\L_{UV}^2$ (with dimensionless $G$)
is assumed to arise from physics at
a new ultraviolet scale $\L_{UV}$.  In order to meet flavor constraints,
$\L_{UV}$ should be much higher than the hypercolor scale $\L_{HC}$
\cite{RC,BCS,PW}.  We calculated the matrix element of $Qqq$
between the vacuum and a chimera state, finding a very small value.
Demanding the hierarchy $\L_{UV}\gg\L_{HC}$, our result implies
that the top quark can only receive a mass which is smaller
by several orders of magnitude than its actual mass.
Conversely, if we insist that the top quark receive the correct mass,
then $\L_{UV}$ would have to be {\em smaller} than $\L_{HC}$.
Either way, the prototype model is ruled out as a realistic
composite-Higgs model.

The remedy might come from enlarging the fermion content of the
hypercolor theory, driving it close to the conformal window, or even into it.
Infrared conformal theories, as well as walking theories,
have anomalous dimensions that (a) can be large,
and (b) can stay roughly constant over many energy decades.
In particular, a large anomalous dimension for the chimera operator $Qqq$
could enhance the 4-fermion coupling $G$ by several orders of magnitude,
eventually leading to a phenomenologically acceptable partially-composite top.

\begin{figure}[t]
\begin{center}
\includegraphics[width=.5\columnwidth]{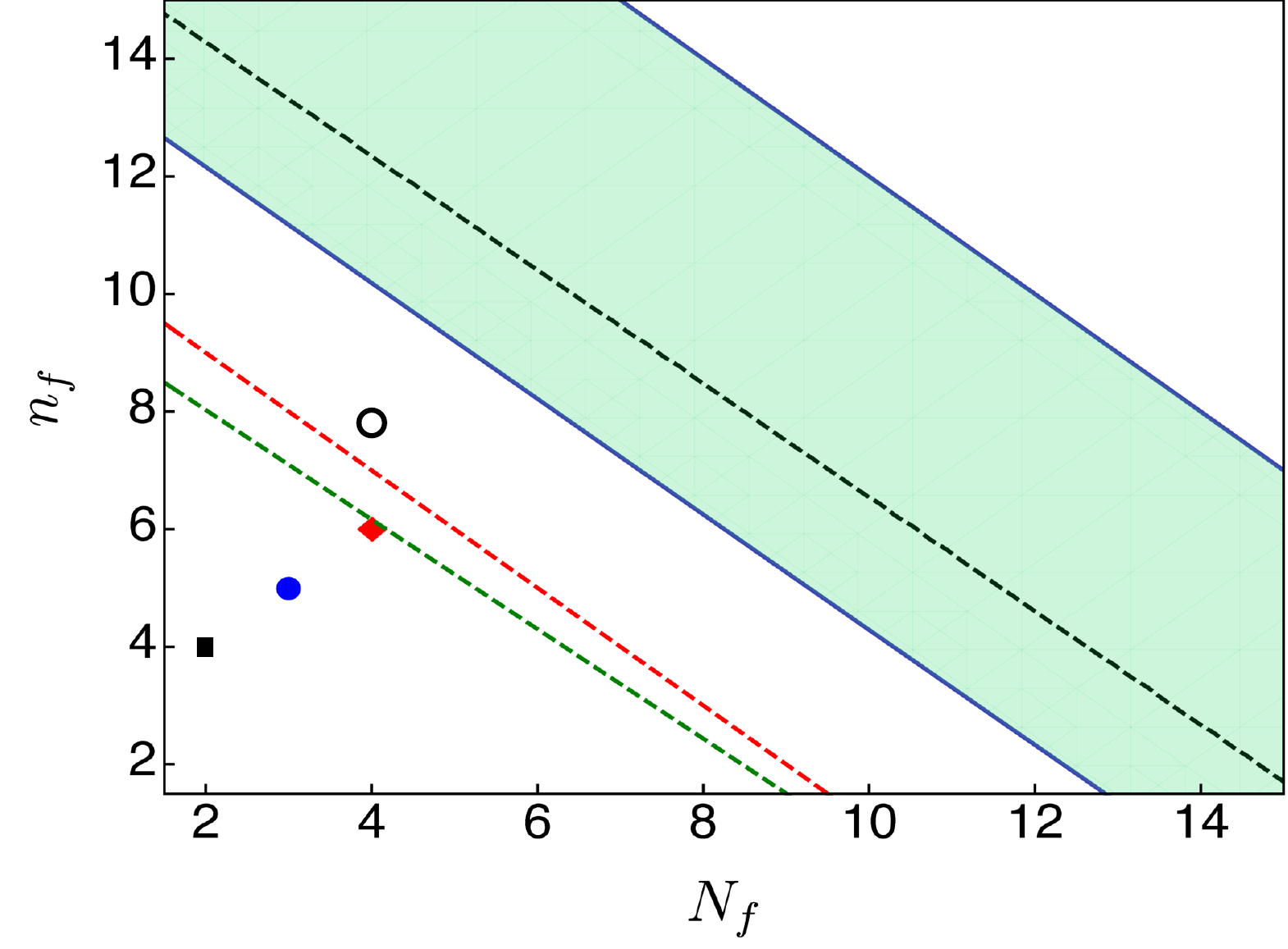}
\end{center}
\vspace*{-2ex}
\caption{
\label{CW}
Estimates for the location of the conformal window
of the SU(4) gauge theory with $N_f$ Dirac fermions in the fundamental
representation and $n_f$ Majorana fermions in the sextet representation.
The uppermost line is the limit of asymptotic freedom.  The green band
represents the analytical estimate of Ref.~\cite{KHL} for the conformal window,
while the dashed lines are other analytical estimates of the bottom
of the window.  Black square: 2+2 model; Blue circle: M6 model;
red diamond: M11 model; open circle: 4+4 model. Adapted from Ref.~\cite{KHL}.}
\end{figure}

Figure~\ref{CW}, adapted from Ref.~\cite{KHL},
shows various analytical estimates
for the conformal window for the SU(4) gauge theory
in the fundamental--sextet plane.  Our prototype ``$2+2$ model''
(with 2 Dirac fermions in each of the fundamental and sextet representations)
is indicated by the black square near the lower-left corner.
Doubling the number of fermions in each representation brings us
to the open circle, or ``$4+4$ model.''  One can see that
the $2+2$ model lies well below the sill of the conformal window
according to all analytical estimates.  By contrast, the $4+4$ model
would be inside the conformal window according to some estimates,
or slightly below it according to the others.
In addition, two models from the Ferretti-Karateev list:
the M6 model mentioned before, as well as the M11 model, can be reached
from the $4+4$ model by giving some of the fermions a large mass.

In summary, the $4+4$ model can be infrared conformal or near conformal
according to analytical estimates, making it a promising candidate for
a composite Higgs model.  It would be interesting to study the $4+4$ model
using similar methods to those we have employed for the $2+2$ model.

\vspace{3ex}
\noindent
{\bf Acknowledgements}
\vspace{1ex}

Our calculations of staggered fermion propagators and currents
were carried out with code derived from version 7.8 of the publicly available
code of the MILC collaboration \cite{MILC}.
Computations for this work were carried out with resources provided
by the USQCD Collaboration, which is funded
by the Office of Science of the U.S.\ Department of Energy.
This material is based upon work supported by the U.S.\ Department of Energy,
Office of Science, Office of High Energy Physics, under Awards No.
DE-SC0010005 (Colorado) and DE-SC0013682 (SFSU),
and by the Israel Science Foundation under grant No.~491/17 (Tel Aviv).
Fermilab is operated by the Fermi Research Alliance, LLC under contract
No.~DE-AC02-07CH11359 with the U.S.\ Department of Energy.


\end{document}